\documentclass[epsf,usegraphicx]{mn2e}

\usepackage{xspace}
\usepackage{amsfonts}
\usepackage{pifont}

\title[AF Psc as seen in K2 data]
{The M4.5V flare star AF Psc as seen in K2 engineering data}

\author[]
{Gavin Ramsay$^{1}$, J. Gerry Doyle$^{1}$ \\
$^{1}$Armagh Observatory, College Hill, Armagh, BT61 9DG, UK\\
}

\date{Accepted 2014 May 28.  Received 2014 May 6; in original form 2014 April 9}

\begin{document}
\outer\def\gtae {$\buildrel {\lower3pt\hbox{$>$}} \over 
{\lower2pt\hbox{$\sim$}} $}
\outer\def\ltae {$\buildrel {\lower3pt\hbox{$<$}} \over 
{\lower2pt\hbox{$\sim$}} $}
\newcommand{\ergscm} {ergs s$^{-1}$ cm$^{-2}$}
\newcommand{\ergss} {ergs s$^{-1}$}
\newcommand{\ergsd} {ergs s$^{-1}$ $d^{2}_{100}$}
\newcommand{\pcmsq} {cm$^{-2}$}
\newcommand{\ros} {\sl ROSAT}
\newcommand{\chan} {\sl Chandra}
\newcommand{\xmm} {\sl XMM-Newton}
\newcommand{\kep} {\sl Kepler}
\def\rchi{{${\chi}_{\nu}^{2}$}}
\newcommand{\Msun} {$M_{\odot}$}
\newcommand{\Mwd} {$M_{wd}$}
\newcommand{\Lsol} {$L_{\odot}$}
\def\Mdot{\hbox{$\dot M$}}
\def\mdot{\hbox{$\dot m$}}
\newcommand{\teff}{\ensuremath{T_{\mathrm{eff}}}\xspace}
\newcommand{\tickYes}{\checkmark}
\newcommand{\tickNo}{\hspace{1pt}\ding{55}}
\newcommand{\src} {KIC 5474065}
\newcommand{\srctwo} {KIC 9726699}

\maketitle

\begin{abstract}
We present the light curve of the little studied flare star AF Psc
(M4.5V) obtained using engineering data from the K2 mission. Data were
obtained in Long Cadence mode giving an effective exposure of 29 min
and nearly 9 d of coverage. A clear modulation on a period of 1.08 d
was seen which is the signature of the stellar rotation period. We
identify 14 flares in the light curve, with the most luminous flares
apparently coming from the same active region. We compare the flare
characteristics of AF Psc to two M4V flare stars studied using
{\kep} data. The K2 mission, if given approval, will present a
unique opportunity to study the rotation and flare properties of late
type dwarf stars with different ages and mass.
\end{abstract}

\begin{keywords}
Physical data and processes: magnetic reconnection -- stars: activity
-- Stars: flares -- stars: late-type -- stars: individual: AF Psc,
{\src}, {\srctwo}
\end{keywords}

\section{Introduction}

NASA's {\kep} mission was launched in March 2009 and spent the next 4
years making near continuous flux measurements of over 160,000 stars
in an area of sky covering 115 square degrees in the constellations of
Cygnus and Lyra (Borucki et al. 2010). Although the prime science
driver for the mission was the discovery of Earth sized planets around
Solar type stars, it provided a wealth of information on objects as
diverse as pulsating stars, accreting systems, supernovae and flare
stars. With the loss of two of the satellites four reaction wheels,
the mission has now evolved into the K2 mission (Howell et
al. 2014). Funding permitting, this will result in a series of
pointings along the ecliptic plane, each lasting $\sim$75 days.  In
the planning stage for the K2 mission, several engineering tests are
being made. Data from the Feb 2014 tests have recently been made
publically available.

The pointing of the original {\kep} was such that the pointing
accuracy was much better than the pixel scale on a timescale much
shorter than the three month quarters (Koch et al. 2010). In the K2
mission, by pointing at fields in the ecliptic plane, photon pressure
from the Sun acts as the only source of force and the two remaining
reaction wheels remove the build up of angular momentum. This causes
the stars to shift by measureable amounts on the CCD
detectors. However, the {\kep} team found that K2 gives photometry
which is within a factor of 2--4 of the original {\kep} data (see
Howell et al 2014).

The almost continuous light curves which {\kep} is able to provide
makes it ideal for the investigation of many types of phenomena
including stellar flares. For instance, Balona (2012) reported
  observations of flares from stars with early A and F spectral types,
  while Maehara et al. (2012) and Shibayama et al. (2013) report
  `super-flares' from Solar type stars. At lower masses, Walkowicz et
al (2011) made a study of flares from cool stars and Gizis et
al. (2013) reported flares from an L dwarf. In Ramsay et al. (2013),
we reported observations of two M4V stars which showed intense flare
activity. Here we report on observations derived from K2 engineering
data on another flare star, AF Psc.

\section{AF Psc}

AF Psc was discovered as a high amplitude ($>$6 mag) flare star nearly
40 years ago (Bond 1976). It was included in a large spectroscopic
sample of nearby M dwarfs and a spectral type of M4.5V was determined
and a distance of 11 pc derived by parallax measurements (Riaz, Gizis
\& Harvin 2006). AF Psc is around the same brightness ($V$=14.4) as
KIC 9726699 ($g$=13.9) but much brighter than KIC 5474065 ($V$=18.1)
both of which were studied by Ramsay et al (2013). It does not appear
to have been the subject of a previous dedicated optical photometric
study, although several flares in the UV were observed on AF Psc using
Galex (Welsh et al. 2006). These authors also reported the detection
of oscillations on a timescale of $\sim$30 sec during flares, which
they interpreted as being due to `slow sausage mode' waves.

\begin{figure*}
\begin{center}
\setlength{\unitlength}{1cm}
\begin{picture}(16,11)
\put(-0.5,0){\includegraphics{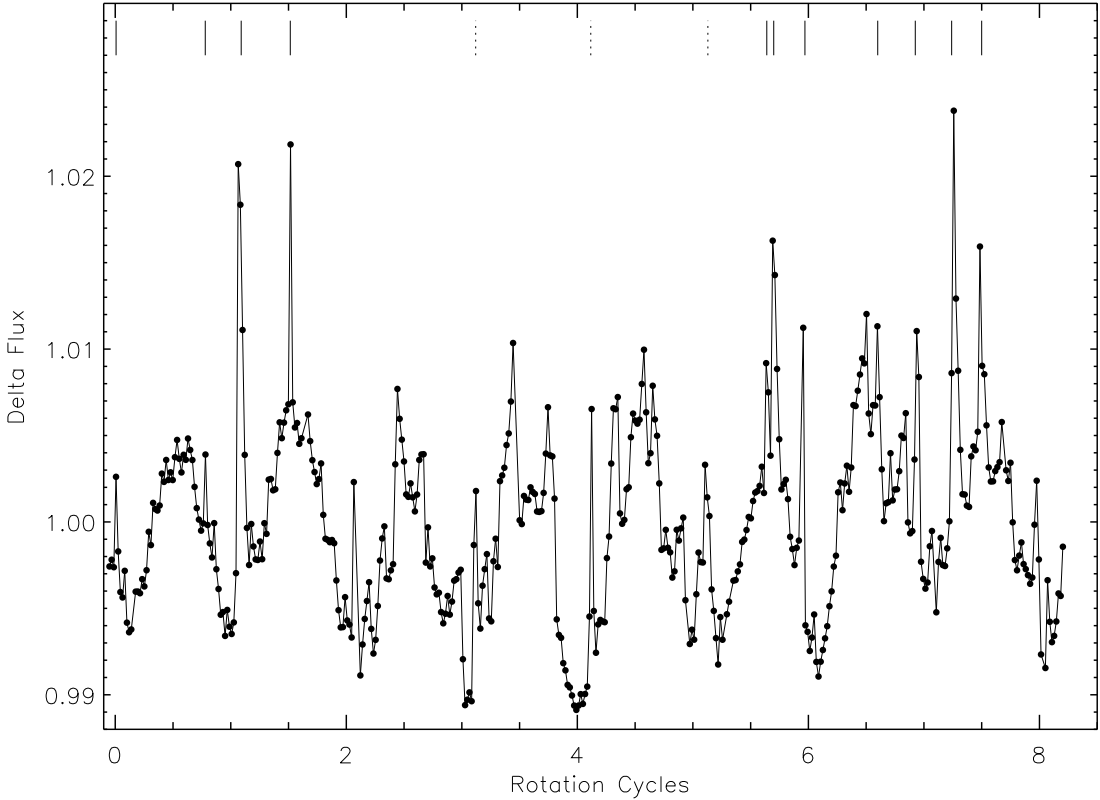}}
\end{picture}
\end{center}
\caption{The light curve of AF Psc made using K2 in engineering data
  where each point has an effective exposure of 29 min. The data has
  been phased ($\phi$=0.0 is defined by minimum flux) on the 1.08 day
  period which is clearly present in the data. The vertical lines
    at the top of the panel note the time of flares in the light
    curve. The three consecutive flares occuring at $\phi\sim$0.12 are
    noted by dashed vertical lines.}
\label{light}
\end{figure*}

\section{K2 Data}

The detector on board {\kep} is a shutterless photometer using 6 sec
integrations and a 0.5 sec readout. The observations of AF Psc were
made in {\it long cadence}, where 270 integrations are summed for an
effective 29.4 min exposure. This contrasts with the observations made
of the two M4V targets KIC 5474065 and KIC 9726699 (Ramsay et al 2013)
which were made in {\it short cadence} where the effective exposure is
58.8 sec.

Observations were carried out in engineering tests from MJD 56692.57
to 56701.50 (2014 Feb 4th to Feb 13). The coverage was therefore 8.9
days in duration. During this time interval there were frequent
thrusts of the spacecraft to ensure pointing accuracy with one large
shift occuring on MJD 56694.86 (or 2.3 days into the time series).

A 50$\times$50 pixel array is downloaded from the satellite for each
target.  To extract a light curve of AF Psc we used the {\tt PyKe}
software (Still \& Barclay
2012){\footnote{http://keplergo.arc.nasa.gov/PyKE.shtml}} which was
developed for the {\sl Kepler} mission by the Guest Observer Office.
We experimented by extracting data from a series of different
combinations of pixels. We found that a mask centered on AF Psc, but
including a relatively faint (USNO-B1 gives $R\sim$18.9) nearby
($\sim$20 arcsec) star consisting of 140 pixels gave the optimal
results. If a smaller number of pixels are used we find that there are
small discontinuities present in the light curve which is the result
of small shifts in the position of the stellar profile over the CCD.
We also experimented with subtracting the background (which increased
in a nearly linear fashion over the course of the observations) in
different ways. We found that using the median value of each time
point to represent the background gave the best results. Finally we
removed time points which were not flagged {\tt `SAP\_QUALITY==0'}
(for instance times of enhanced solar activity).

\section{Results}

We normalised the extracted light curves by dividing the data by the
mean background subtracted flux.  The light curve of AF Psc shows a
clear modulation on a period of 1.08$\pm$0.08 days (Figure
\ref{light}). Given AF Psc is an M4.5V star, this modulation is caused
by the rotation of spots or active regions on the photosphere.  The
first few rotation cycles have smooth profiles, but then become more
complex (double horn shaped at maximum) which indicates that active
regions appear and dissappear on relatively short timescales.

In Ramsay et al. (2013), we used an automatic routine to identify
flares in the {\kep} data of two M4V stars. Given the quality of the
present light curve is relatively lower (and the time coverage much
shorter) we decided to manually identify flares in the light curve of
AF Psc. We searched for events which showed a rapid rise in flux and an
exponential decline which is characteristic of stellar flares. Given
each exposure is 29 min, very short duration flares are likely to
either be missed completely or seen as a flux increase in only one
time point.  We were rather conservative in identifying points as
flares and did not (for instance) flag the enhanced flux point at 2.05
rotation cycles (Figure \ref{light}) as this coincided with a
significant shift in the stellar profiles. In some cases it was rather
subjective whether an event was a flare or not -- for instance we did
not identify features in rotation cycle 2 (Figure 1) as flares, but
rather the general variation seen in the rotation curves of active
stars. We identified 14 flares in AF Psc (which are marked in Figure
1) over the whole 9 days of data.

\begin{figure}
\begin{center}
\setlength{\unitlength}{1cm}
\begin{picture}(8,12)
\put(-0.5,-0.5){\includegraphics{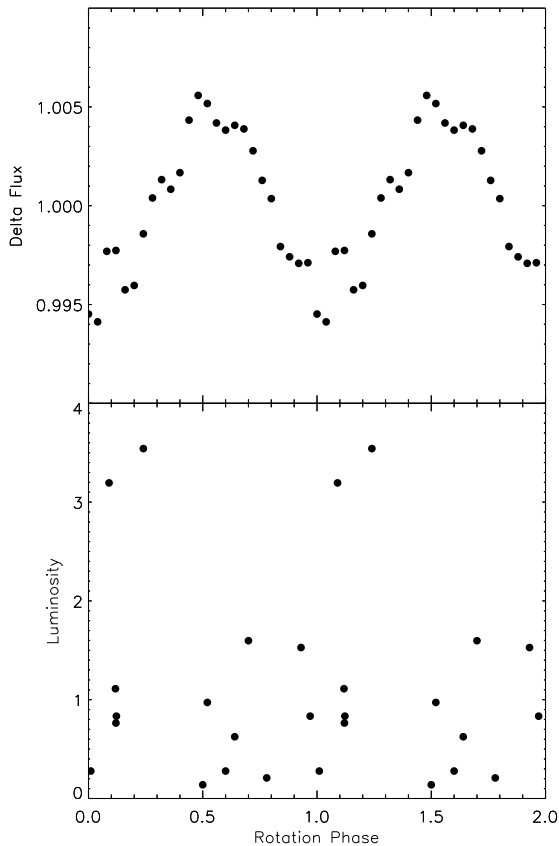}}
\end{picture}
\end{center}
\caption{Top Panel: The light curve of AF Psc phased and binned on the rotation
  period of 1.08 days. Lower Panel: The energy of the flares plotted as a 
function of rotation phase.}
\label{fold}
\end{figure}

To determine the luminosity of the flares we assume that a star with
spectral type M4.5V has an luminosity $L=2.5\times10^{30}$ erg
s$^{-1}$ (see Ramsay et al 2013 for details). We then sum up the area
in the flare assuming this base luminosity. Given we do not use any
model for the flare profile and that the time resolution is rather
low, there is some degree of uncertainty in the estimate for each
individual flare. However, it does indicate the general
characteristics of the flare behaviour of this source.  We find that
the flares seen in AF Psc have a luminosity in the range
$\sim0.1-3.0\times10^{32}$ erg in the {\kep} band-pass.  This compares
to $L=1.1-7.3\times10^{32}$ ergs for KIC 5474065 and
$L=0.01-2.2\times10^{32}$ ergs for KIC 9726699.

We show the folded and binned light curve of AF Psc in Figure
\ref{fold} together with the rotation phase and luminosity of each
flare. We note that the most luminous flares were seen between
$\phi$=0.1--0.3. Indeed, in Figure 1 it is seen that there are three
consecutive rotation cycles where a flare is seen at $\phi\sim$0.12.
This indicates that there is a region on the star which is active over
more than one rotation cycle.

\begin{figure}
\begin{center}
\setlength{\unitlength}{1cm}
\begin{picture}(8,6)
\put(-0.8,-0.2){\includegraphics{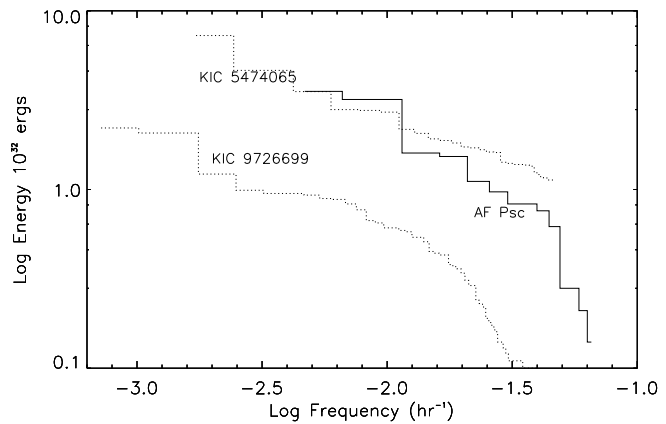}}
\end{picture}
\end{center}
\caption{The cumulative energy distribution of flares (in the {\kep}
  band-pass) as seen in AF Psc, {\src} and KIC 9726699.}
\label{rate}
\end{figure}

We show the cumulative flare-frequency distribution in Figure
\ref{rate} for AF Psc and KIC 5474065 and KIC 9726699. Interestingly,
despite being a more rapid rotator than KIC 5474065 (1.08 d compared
to 2.47 d), the flare-frequency distribution of these sources are very
similar. The distribution of AF Psc goes to lower luminosities since
it is a much brighter source (and hence less luminous flares may have
been missed in KIC 5474065). In contrast, KIC 9726699, although having
a very similar spectral type (M4V) is a more rapid rotator (0.59 d).
AF Psc does not show the high amplitude flares seen in KIC 5474065 but
this maybe due to the shorter timeline of the data (8.9 d compared to
24.2 d) and the longer exposure time for each photometric point.

\section{Conclusions}

If the K2 mission proceeds as hoped, it will cover a number of
clusters (Howell et al 2014) which have ages ranging from the very
young (Taurus-Auriga Association at 2 Myr), to the not-so-young (the
Pleiades at 125 Myr) to the positively old (M67 at 3.6 Gyr). The
engineering data presented here of AF Psc demonstrate that K2 has the
photometric accuracy to identify the rotation period and flare rate of
M dwarf stars even in long cadence mode and over a time interval
significantly shorter than that planned for K2 in full operation. K2
will give an unique opportunity to determine how the stellar rotation
period and flare rate of late type dwarfs are effected by age, mass
and metallicity.  West et al 2008) showed that a marked change in
activity levels occurs around the spectral type M4. K2 will allow the
the determination of key observables for dozens of stars with spectral
type in the range M0--M8 and hence address the underlying physical
reasons for this.

\section{Acknowledgements}

This paper includes data collected by the Kepler spacecraft using
2-wheel mode. Funding for the Kepler spacecraft is provided by the
NASA Science Mission Directorate. Some of the data presented in this
paper were obtained from the Mikulski Archive for Space Telescopes
(MAST). STScI is operated by the Association of Universities for
Research in Astronomy, Inc., under NASA contract NAS5-26555. Support
for MAST for non-HST data is provided by the NASA Office of Space
Science via grant NNX09AF08G and by other grants and contracts. This
work made use of PyKE, a software package for the reduction and
analysis of Kepler data. This open source software project is
developed and distributed by the NASA Kepler Guest Observer Office. We
thank Tom Barclay, Martin Still and Steve Howell for useful
advice. Armagh Observatory is supported by the Northern Ireland
Government through the Dept Culture, Arts and Leisure.

\end{document}